\documentclass[conference, 10pt]{IEEEtran}

\usepackage{cite}
\usepackage{url}
\usepackage{graphicx}
\usepackage{color}
\usepackage{placeins}
\usepackage{float}
\usepackage{tabularx,colortbl}
\usepackage{ifthen}
\usepackage{csquotes}

\usepackage{fancyhdr}
\fancypagestyle{firstpage}{
  \fancyhf{} 
  \fancyfoot[C]{} 
}
\usepackage{caption}
\usepackage{tikz}
\usepackage{subcaption}
\usepackage{graphicx}

\usetikzlibrary{
    shapes.geometric,
    shapes.misc,
    arrows.meta,
    positioning,
    calc,
    fit,
    backgrounds,
    decorations.pathreplacing,
    patterns
}

\makeatletter

\newcounter{author}
\renewcommand{\author}[2][]{
   \stepcounter{author}
   \@namedef{author@\theauthor}{#2}
   \@namedef{authorlabel@\theauthor}{#1}
}

\newcounter{address}
\newcommand{\address}[2][]{
   \stepcounter{address}
   \@namedef{address@\theaddress}{#2}
   \@namedef{addresslabel@\theaddress}{#1}
}

\newcommand{\alsep}{and}

\def\newmaketitle{\par%
  \begingroup%
  \normalfont%
  \def\thefootnote{}
  \def\footnotemark{}
  \let\@makefnmark\relax
  \footnotesize
  \footnotesep 0.7\baselineskip
  \normalsize%
  \twocolumn[\thenewmaketitle\@IEEEaftertitletext]%
  \if@IEEEusingpubid
     \enlargethispage{-\@IEEEpubidpullup}%
  \fi
  \endgroup
  \setcounter{footnote}{0}\let\maketitle\relax\let\@maketitle\relax
  \gdef\@thanks{}%
  \let\thanks\relax}

\def\thenewmaketitle{
  \newpage
  \begin{center}%
    \vskip0.2em{\Huge\@IEEEcompsoconly{\sffamily}\@IEEEcompsocconfonly{\normalfont\normalsize\vskip 2\@IEEEnormalsizeunitybaselineskip
   \bfseries\large}\@title\par}\vskip1.0em\par%
    \vspace{1ex}
    \newcounter{c@author}
    \newcounter{c@tmp}
    \ifthenelse{\value{author}=2}{%
      \newcommand{\liand}{ and }}{%
      \newcommand{\liand}{, and }}
    \ifthenelse{\value{address}<2}{%
      \@nameuse{author@1}%
      \stepcounter{c@author}%
      \whiledo{\value{c@author}<\value{author}}{%
        \setcounter{c@tmp}{\value{author}}%
        \addtocounter{c@tmp}{-\value{c@author}}%
        \ifthenelse{\value{c@tmp}=1}{%
          \renewcommand{\alsep}{\liand}}{\renewcommand{\alsep}{, }}%
        \stepcounter{c@author}\alsep \@nameuse{author@\thec@author}}\\%
    }
    {
      \@nameuse{author@1}${}^{(\ref{\@nameuse{authorlabel@1}})}$%
      \stepcounter{c@author}%
      \whiledo{\value{c@author}<\value{author}}{%
      \setcounter{c@tmp}{\value{author}}%
      \addtocounter{c@tmp}{-\value{c@author}}%
      \ifthenelse{\value{c@tmp}=1}{%
        \renewcommand{\alsep}{\liand}}{\renewcommand{\alsep}{, }}%
      \stepcounter{c@author}\alsep \@nameuse{author@\thec@author}%
        ${}^{(\ref{\@nameuse{authorlabel@\thec@author}})}$%
      }
    }
    \vspace{0.2ex}

    \ifthenelse{\value{address}>0}{%
      \ifthenelse{\value{address}=1}{
        {\@nameuse{address@1}}
      }
      {
        \newcounter{c@address}

        \begin{center}
        \whiledo{\value{c@address}<\value{address}}
        {
          \refstepcounter{c@address}
            ${}^{(\thec@address)}$\,%
              \label{\@nameuse{addresslabel@\thec@address}}%
              \@nameuse{address@\thec@address}\\ %
        }
        \end{center}
      } 
    }
    {
      \relax
    }
  \end{center}
}

\makeatother

\title{Agentic Physical-AI for Self-Aware RF Systems}

\author[org1]{Linuka Ratnayake}
\author[org1]{Danidu Dabare}
\author[org1]{Sanuja Rupasinghe}
\author[org1]{Warren Jayakumar}
\author[org2]{\protect\\Dileepa Marasinghe}
\author[org1]{Chamira U. S. Edussooriya}
\author[org3]{Arjuna Madanayake}

\address[org1]{University of Moratuwa, Moratuwa, Sri Lanka (chamira@uom.lk)}
\address[org2]{University of Oulu, Oulu, Finland (dileepa.marasinghe@oulu.fi)}
\address[org3]{Florida International University, Miami, FL, USA (amadanay@fiu.edu)}

\begin{document}

\newmaketitle
\thispagestyle{firstpage}
\begin{abstract}
Intelligent control of RF transceivers adapting to dynamic operational conditions is essential in the modern and future communication systems. We propose a multi-agent neurosymbolic AI system, where AI agents are assigned for circuit components. Agents have an internal model and a corresponding control algorithm as its constituents. Modeling of the IF amplifier shows promising results, where the same approach can be extended to all the components, thus creating a fully intelligent RF system.


\end{abstract}
\vspace{-0.5em}

\section{Introduction}
Artificial intelligence (AI) and machine learning (ML) have been reshaping wireless and radio frequency (RF) systems in recent years. Data-driven techniques such as neural receivers are being explored enabling improved capability to handle channel non-linearities and hardware impairments compared to conventional architectures. Reinforcement learning approaches, exemplified by IARPA’s End-to-End Generative Waveforms (EndGen), are enabling clean-slate waveform, modulation, and coding design, while AI based circuit design methods learn from existing designs and data to generate new hardware solutions~\cite{karahan2024deep,Lai2025}. In this paper, we propose a multi-agent RF transceiver architecture, inspired by prior cognitive RF systems \cite{cousik2019cogrfnewfrontiermachine,mapcon2025}, that dynamically adapts at runtime to varying operational conditions. In the proposed framework, each component is associated with an AI agent and characterized by a set of tunable parameters. These parameters are embedded within carefully selected data-driven circuit models that form the basis of digital twins, providing a physics based simulation environment. The digital twins are neurosymbolic in nature, i.e., RF behavior may be modeled using a physics based formulations, data-driven ML models or a combination of both. The resulting multi-agent system continuously self adapts, optimizes, and learns from real time observations of the RF environment, ensuring robust operation across varying and unforeseen conditions.

\section{Multi-agent RF Transceiver Architecture}

We propose a generalized architectural framework for controlling RF systems where hardware components are abstracted as multiple autonomous neurosymbolic agents. This framework is agnostic to the specific RF front-ends and can be scaled to complex transceivers, enabling different hardware components to use a standardized interface and work together. This software layer will provide control functionality to the physical hardware circuit, through a multi-agent AI framework that operates on an abstraction layer above the physical hardware.

\begin{figure}[t!]
    \centering
    \includegraphics[width=\linewidth]{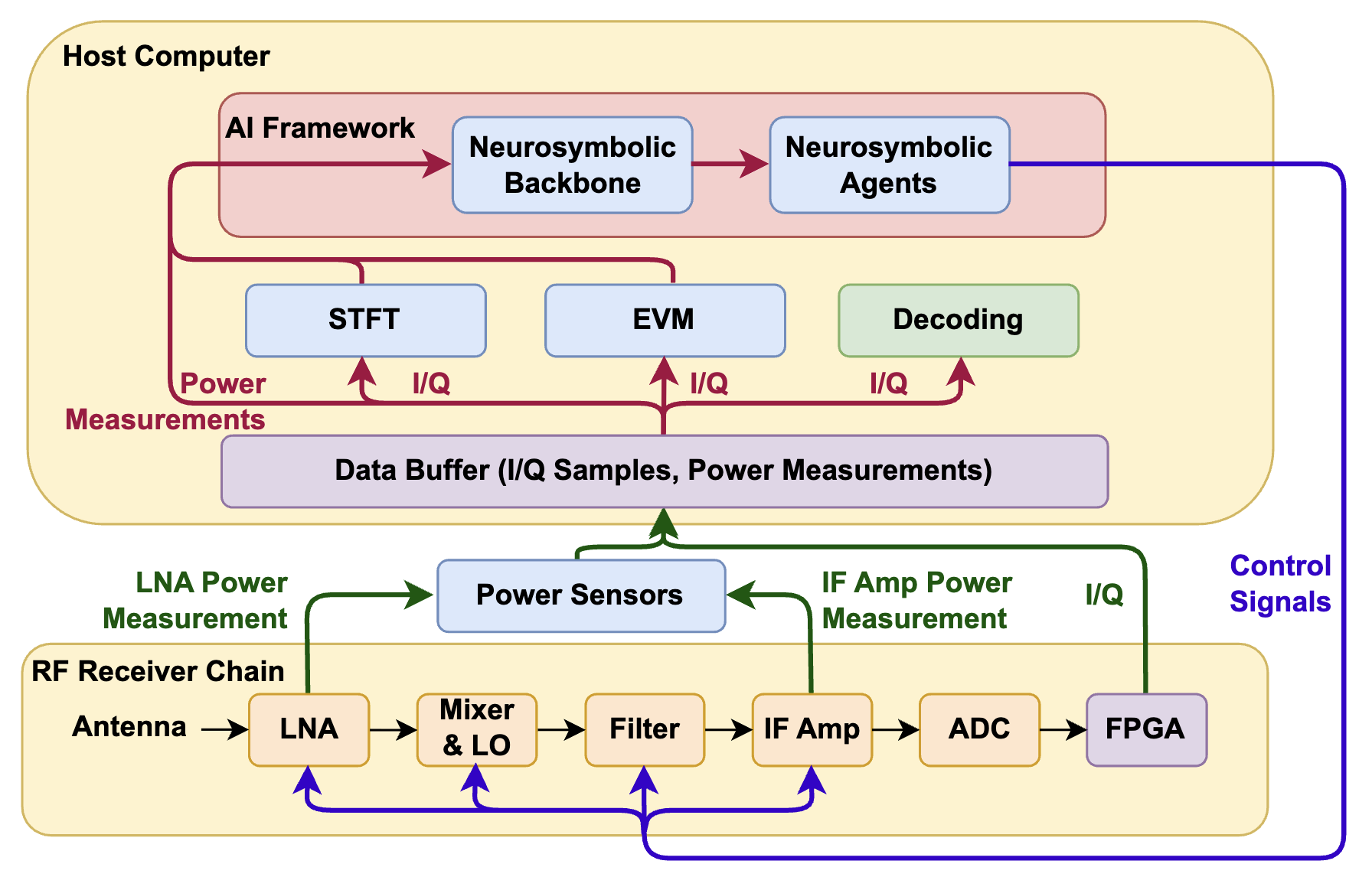}
    \caption{System architecture shows that the multi-agent AI framework optimizes the physical RF receiver using a neurosymbolic architecture driven by real-time signal features (STFT, EVM) and sensor feedback.}
    \label{fig:architecture}
    \vspace{-4ex}
\end{figure}

For each abstracted component in the physical circuit such as low noise amplifiers (LNAs), mixer, filters, and intermediate frequency (IF) amplifiers, there is an agent associated with that particular hardware component, as shown in Fig.~\ref{fig:architecture}. Each agent consists of two primary internal components. The first is a forward model of the associated hardware component's behavior. This model is neurosymbolic in nature, i.e., it uses rule based approaches, data based approaches, or a combination of both. It provides an inexpensive surrogate for the physical component, reducing the need for extensive hardware experimentation or computationally expensive simulations. The second component is the control algorithm for the hardware block, which may likewise employ neurosymbolic control mechanisms. As an example, amplifiers are modeled using an augmented real valued time delay neural network (ARVTDNN) \cite{arvtdnn} which can model memory effects and non-linearities. Furthermore, filters are represented using physics based symbolic models derived from transfer functions and other components may be modeled with similar techniques. 

The forward models of individual components together form a digital twin of the system, enabling advanced simulation of system behavior. This digital twin can be used to provide various operating scenarios and observe how the system processes the signal. The digital twin can also be used to generate a large dataset, on which the agents can be trained on. In this case, a two-stage approach is used such that, input signal power and tunable component parameters are randomly varied to observe the resulting performance, creating a broad library of operating points first. Then a hybrid Bayesian optimization approach is applied to explore the dataset and iteratively guides the search toward parameter settings that maximize the performance. This combined strategy produces a comprehensive dataset used to train the supervised learning models in the control algorithms.

To demonstrate the multi-agent framework, we present an example setup, that will be realized using a conventional RF receiver (Rx) chain, see, Fig.~\ref{fig:architecture} for the overall system architecture. Here, all components in the Rx chain are tunable, i.e., parameters such as LNA supply voltage, local oscillator frequency and amplitude, filter bandwidth, and IF amplifier gain can be dynamically adjusted. The purpose of the control algorithms inside the agents is to predict these dynamic configurations.

The circuit is probed after the LNA and after the IF amplifier, to measure signal power which will give crucial insights to the AI framework about the signal. Following the analog to digital converter (ADC), digital signal processing is carried out to extract in-phase and quadrature (I/Q) samples, that will be used to
compute the short time Fourier transform (STFT) of the received signal and the error vector magnitude (EVM), which captures the overall signal quality at the end of the receiver chain. These features together with the two power measurements are provided as inputs to the AI framework for decision making. The goal is to minimize the EVM of the received signal. The multi-agent AI framework takes various observations and measurements of the real-time system, and the individual agents use their internal models to simulate, and collectively decide the optimal configuration for each hardware component, which will proactively change the hardware configurations, to achieve optimal output.


\section{Preliminary Results}

\begin{figure}[t!]
    \centering
    \includegraphics[width=0.9\linewidth]{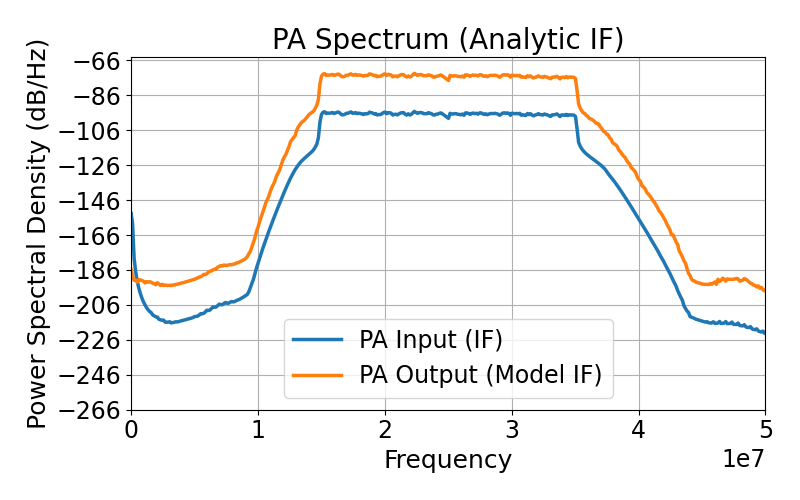}
    \vspace{-1ex}
    \caption{Input and output power spectral densities shows that the ARVTDNN model accurately replicates the frequency response of the IF amplifier, validating the digital twin.}
    \label{fig:ARVTDNN_spectrum}
    \vspace{-3ex}
\end{figure}
\begin{figure}[t!]
    \centering
    \includegraphics[width=0.9\linewidth]{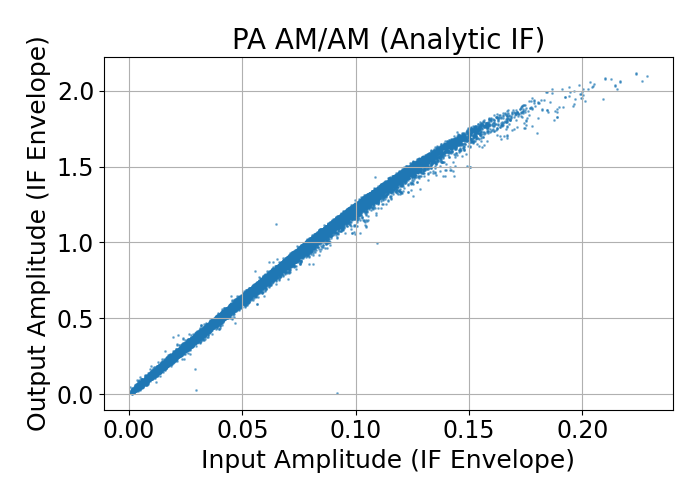}
    \vspace{-1ex}
    \caption{Scatter plot of AM/AM compression of the input vs. output envelope amplitude shows that the model successfully captures the IF amplifier's nonlinear gain characteristics and memory effects.}
    \label{fig:ARVTDNN_compression}
    \vspace{-3ex}
\end{figure}

Fig.~\ref{fig:ARVTDNN_spectrum} presents the actual and predicted behavior of the ARVTDNN model developed for the IF amplifier (LMH6401), demonstrating that the frequency response is modeled with high accuracy. Furthermore, the corresponding amplitude characteristics are presented in Fig. \ref{fig:ARVTDNN_compression}. We note that the ARVTDNN model exhibits a high degree of fidelity to the measured hardware behavior. Beyond learning the linear gain characteristics, the model successfully captures nonlinear distortions and memory effects inherent in the amplifier. This serves as a reliable surrogate for the physical hardware, enabling accurate offline optimization of control parameters with strong confidence in their transferability to real world deployment. 


Similarly, all other components can be modeled using neurosymbolic approaches with high fidelity, thus creating the complete and well matched digital twin, providing highly reliable method of simulation and dataset generation.

\section{Discussion and Conclusion}

The proposed multi-agent architecture provides reliable and real-time control functionality to the physical RF hardware. By modeling hardware blocks as individual agents, the system enables decentralized control and scales well with hardware changes, supporting the \enquote{Physical AI} paradigm by tightly coupling digital twins with real RF behavior.

An advantage is real-time operation on dynamic conditions with proactive actions. This capability is critical for 6G, where diverse frequencies, bandwidths, and protocols must coexist, enabling a self-aware transceiver to replace multiple hardware-specific implementations. The main trade-off is the computational overhead of continuous AI inference, which may limit use in energy-constrained edge devices but is well suited to infrastructure-scale systems such as base stations. 


%
\bibliographystyle{ieeetr}
\bibliography{bibliography}


\end{document}